\documentclass[prl,aps,twocolumn,showpacs]{revtex4}
\usepackage{amsmath}
\usepackage{amsfonts}
\usepackage{epsfig}
\usepackage{bm}

\begin{document}

\title{Entanglement entropy of critical spin liquids}

\author{Yi Zhang, Tarun Grover and  Ashvin Vishwanath}

\affiliation{Department of Physics, University of California, Berkeley, California 94720, USA}

\begin{abstract}
Quantum spin liquids are phases of matter whose internal structure is not captured by a local order parameter. Particularly intriguing are critical spin liquids, where strongly interacting excitations control low energy
 properties. Here we calculate their bipartite entanglement entropy that characterize their quantum structure. In particular we calculate  the Renyi entropy $S_2$, on model wavefunctions obtained by
 Gutzwiller projection of a Fermi sea. Although the wavefunctions are not sign positive, $S_2$ can be calculated on relatively large systems ($>$324 spins), using the variational Monte Carlo technique.
 On the triangular lattice we find that entanglement entropy of the projected Fermi-sea state  violates  the boundary law, with $S_{2}$ enhanced by a logarithmic factor. This is an unusual result for
 a bosonic wave-function reflecting the presence of emergent fermions. These techniques can be extended to study a wide class of other phases.

\end{abstract}

\pacs{}

\maketitle

Entanglement properties of a ground state wave-function can serve as a diagnostic for characterizing a phase of matter \cite{eisert2010}. Such a characterization is especially useful when a phase does not break
any symmetry and consequently does not offer a local order parameter. Two examples of this observation are the presence of a non-zero topological entanglement entropy in a fully gapped topological ordered state
\cite{kitaev2006} and the relation between the edge state spectrum and the entanglement spectrum for the quantum Hall states \cite{li2008}. Entanglement thus encodes fundamental properties of a quantum phase, and intersects diverse fields including quantum information, many body physics and high energy theory \cite{eisert2010}.

\textit{Critical} spin-liquids are a distinct class of states
that also do not have a local order parameter. These gapless states are less well understood compared to the gapped spin-liquids which have topological order. Interestingly though, a large class of experiments suggest that they
may well have been already realized in certain organic materials \cite{yamashita2010}, which are quantum antiferromagnets on the triangular lattice.  To gain a deeper understanding of these phases, in this paper we investigate
 the quantum structure of a class of critical spin-liquids from an information theoretic point of view. We demonstrate that entanglement properties of model ground states can be calculated, even on fairly large systems
 involving $>324$ spins. In contrast to quantum Monte Carlo\cite{hastings2010}, we deal with non-positive wavefunctions that potentially describe ground states of frustrated quantum magnets.


Both gapped as well as gapless spin-liquids
may be obtained within a slave-particle approach \cite{wen2004}, where one expresses the spin operator in terms of a product of operators ($f$),
 denoting emergent particles : $\vec{S}=f^\dagger_\sigma\frac{\vec{\sigma}_{\sigma\sigma'}}{2}f_{\sigma'}$. In a spin liquid phases, these emergent particles are the appropriate
 low energy excitations. Inevitably, they appear coupled to gauge fields. Gapped spin liquids are described by topological quantum field theories and are relatively well understood.
  In contrast, spin-liquids with gapless, strongly interacting excitations (critical spin liquids), are described by matter-gauge theories which are harder to analyze \cite{gauge_FS}.


 A critical spin liquid, the spinon fermi sea (SFS) state, has been invoked \cite{motrunich05} to account for the intriguing phenomenology of aforementioned triangular lattice organic compounds \cite{yamashita2010}.
 In the SFS state, the $f$ particles hop on the triangular lattice sites giving rise to a Fermi sea, while strongly interacting with an emergent `electromagnetic' $U(1)$ gauge field. The metal like specific heat and thermal conductivity seen
 in these materials is potentially an indication of the spinon Fermi surface. A candidate ground state spin wavefunction for the SFS state suggested by the slave-particle formalism, is obtained by projecting out all doubly occupied states
 from the Fermi sea wavefunction. This Gutzwiller projection technique works well in one dimension. There, the projected Fermi sea spin wavefunction captures long distance properties of the Heisenberg chain, and is even the exact ground state of the Haldane-Sastry \cite{HaldaneSastry} Hamiltonian. Similar rigorous results are not available in two dimensions. However, the Gutzwiller projected Fermi sea is known to have excellent variational energy for
 the $J_2-J_4$ spin model on the triangular lattice, which is believed to be appropriate for the forementioned triangular lattice compounds\cite{motrunich05}. Detailed comparisons between the projected wavefunction and exact numerics have also been made\cite{block2009}.

 Using the Variational Monte-Carlo technique, we calculate the bipartite entanglement entropy, in particular Renyi entropy $S_{2}$ of a critical spin liquid - the conjectured spinon Fermi sea state on the triangular lattice. We find
 a violation of the boundary law,
 with $S_{2}$ enhanced by a logarithmic factor, an unusual  result for a bosonic wavefunction strongly suggesting the presence of emergent fermions with a Fermi surface. This is also consistent with the recent exact numerical studies
 on multi-leg ladder spin-$1/2$ ring exchange models where the central charge is found to increase in proportion to the number of legs \cite{block2009}. We also calculate Renyi entropy for an algebraic spin
 liquid state \cite{wen2002, wen2004} obtained by Gutzwiller projecting the wave function of free Fermions on a $\pi$ flux square lattice.  This state is found to obey the area law, consistent with the presence of
 emergent Dirac  fermions in the system. Finally, we calculate Renyi entropy for projected Fermi sea state on the square lattice. Here a nested Fermi sea is present before projection. In this case the projection
 is found to result in long range magnetic order (see also \cite{li2011}) and a significant reduction in entanglement entropy, especially of the area law violating term, compared to the unprojected Fermi gas. These different trends are found to set in even at relatively small system sizes, suggesting  that this probe may be applied in the context of exact diagonalization and DMRG.

{\em Renyi Entanglement Entropy:} Given a normalized wavefunction, $|\phi\rangle$, and a partition of the system into subsystems $A$ and $B$, one can trace out the subsystem B to give a density matrix on A: $\rho_A = {\rm Tr}_B |\phi\rangle\langle\phi|$.
The Renyi entropies are defined by:
\begin{equation}
S_n = \frac1{1-n}\log({\rm Tr}\rho_A^n)
\end{equation}
It is common to pay special attention to the von Neumann entropy, $S_1= -{\rm Tr}[\rho_A \log \rho_A]$ (obtained by taking the limit $n \rightarrow 1$). However, the Renyi entropies seem
equally good measures of entanglement as they share many properties with von Neumann entropy $S_1$. Here we will focus on $S_2 = -\log\left ({\rm Tr}\rho_A^2 \right )$.
 For ground states of local Hamiltonians, both $S_1$ and $S_2$ are expected to follow a boundary law, i.e. they only grow as the surface area of the boundary of region $A$ \cite{eisert2010}. A well known
violation occurs for critical phases in 1D, where in contrast to a constant as expected from the area law,  $S_1, S_2 \sim  \log L_A$, where $L_A$ is
 the size of system $A$ embedded inside an infinite line\cite{holzhey1994}. In higher dimensions, $D=2,3$ an area law is believed to hold for $S_1, S_2$ even for gapless states such as ordered phases with goldstone modes, and quantum critical points \cite{ryu2006, metlitski09, hastings2010}.
 Both $S_1, S_2$ violate the area law in higher dimensions for free fermions with a Fermi surface \cite{Klich,Wolf} and also share identical topological entanglement entropies for gapped phases\cite{flammia2009}.
 Therefore $S_2$ seems as good a measure of physical properties as the von Neumann entropy, and is often easier to calculate \cite{hastings2010}
as we detail now for our problem. Furthermore we note the inequality  $S_1 \geq S_2$. This will be important when we find violation of area law for $S_2$ as
this would imply area law violation for the von Neumann entropy $S_1$ as well.

{\em Monte Carlo Evaluation:} Consider the wavefunction  $\phi(a,b)$, where $a$ ($b$) be the configuration of subsystem A (B). The Renyi entropy $S_2$ is conveniently expressed by imagining two copies of the system. Construct the product wavefunction: $|\Phi\rangle = \sum_{a,b,a',b'} \phi(a,b)\phi(a',b')|a,b\rangle|a',b'\rangle$, and define the ${ {\rm Swap}}_A$ operator \cite{hastings2010}, which swaps the configurations of the $A$ subsystem in the two copies: ${{\rm Swap}}_A |a,b\rangle|a',b'\rangle = |a',b\rangle|a,b'\rangle$ then it is easily shown:

\begin{equation}
e^{-S_2} =\frac{ \langle \Phi|{{\rm Swap}}_A |\Phi\rangle}{\langle \Phi|\Phi\rangle}
\label{swap}
\end{equation}
This can be rewritten in a form suitable for Monte Carlo evaluation. Defining configurations $\alpha_1=a,\,b$, $\alpha_2=a',\,b'$ and $\beta_1=a',\,b$, $\beta_2=a,\,b'$,
\begin{equation}
\left\langle {\rm {\rm Swap}}_{A}\right\rangle  =  \underset{\alpha_{1}\alpha_{2}}{\sum}\rho_{\alpha_{1}}\rho_{\alpha_{2}}f\left(\alpha_{1},\alpha_{2}\right)
\end{equation}
Here $f\left(\alpha_{1},\alpha_{2}\right)=\frac{\phi_{\beta_1}\phi_{\beta_2}}{\phi_{\alpha_1}\phi_{\alpha_2}}$ is averaged by sampling with the normalized weights $\rho_{\alpha_{i}}= |\phi_{\alpha_i}|^2/{\sum_{\alpha_i}|\phi_{\alpha_i}|^2}$

The Gutzwiller projected wavefunctions
we will consider can be written as products of determinants. Such wavefunctions can be efficiently evaluated, which forms the basis for Variational Monte Carlo technique \cite{gros1989}.
Since the entanglement entropy is at the very least expected to scale as the length of the boundary of subsystem $A$, $\sim L_A$, $\left\langle {\rm Swap}_{A}\right\rangle$ is at least exponentially small in $L_A$. Evaluating this small quantity may pose a problem - one solution called the `ratio trick' discussed in Ref \cite{hastings2010} in the valence bond basis unfortunately does not work in our case, since the relevant states do not always have non-vanishing overlap unlike in that basis.

A more serious problem arises from the fact that $f\left(\alpha_{1},\alpha_{2}\right)$ is not necessarily positive, for the wavefunctions we will study. This is in contrast to Quantum Monte Carlo studies
with positive ground state wavefunctions\cite{hastings2010}. With time reversal symmetry, $f\left(\alpha_{1},\alpha_{2}\right)$ is real but can fluctuate in sign.
The calculational effort for $\left\langle {\rm Swap}_{A}\right\rangle$ is then found to scale exponentially with $L_A$. Although Variational Monte Carlo schemes are generally free of the fermionic sign problem, we encounter a mild variant here. Below we partially mitigate this by introducing a `sign trick' algorithm that separates ${\rm Swap}_{A}$ into a product of two factors, both calculable within
Variational Monte Carlo. This renders the calculation doable
within few percent error for boundary sizes $L_A \lesssim 10$. For positive wavefunctions (i.e. ones that obey Marshall sign)
we find an immense simplification, and the calculational cost only scales polynomially with $L_A$ as in conventional Variational Monte Carlo. For a fixed value  of boundary $L_A$, the cost of
computation always scales polynomially with the total system size $L$ irrespective of the wavefunction.

{\em Sign Trick:} After some algebra \cite{SupplementaryMaterial}, Eqn.\ref{swap} can in general be rewritten as a product of two averages:
\begin{eqnarray*}
\left\langle {\rm Swap}_{A}\right\rangle &=&  \underset{\alpha_{1}\alpha_{2}}{\sum}\rho_{\alpha_{1}}\rho_{\alpha_{2}}|f\left(\alpha_{1},\alpha_{2}\right)|  \left [ \underset{\alpha_{1}\alpha_{2}}{\sum}\tilde{\rho}_{\alpha_{1},\alpha_2}e^{i\phi \left(\alpha_{1},\alpha_{2}\right)}\right ] \\
& = & \langle {\rm Swap}_{A,mod}\rangle \langle {\rm Swap}_{A,sign}\rangle
\end{eqnarray*}

 For real wavefunctions, the phase factor reduces to a sign. Empirically, we find this leads to significant gains, the advantages are detailed in \cite{SupplementaryMaterial}.
We benchmarked our algorithm for three free fermion tight binding problems on: 1) A one dimensional chain of $L = 200$ sites with $L_A$ up to 100 sites, 2) An $18 \times 18$ square lattice with the linear size $L_A$ up to 7 sites. 3) A honeycomb (graphene) lattice with Dirac dispersion. We find very good agreement with the exact results \cite{SupplementaryMaterial} that were calculated using the correlation matrix technique \cite{peschel2009}.

{\em Gutzwiller Projected Spin Liquid Wavefunctions:}
Next we calculate Renyi entropy for the problems of our actual interest namely projected Fermi liquid wave functions which are considered good ansatz for
ground states of critical spin-liquids. We analyze two different classes of critical spin-liquids: states that at the slave-particle mean-field level have a
full Fermi surface of spinons and those with only nodal fermions. For a triangular lattice with uniform hopping $t_{rr'} = t$ one obtains a Fermi surface of spinons at the mean-field level while for a square-lattice with $\pi$
flux through every plaquette (i.e. $\Pi_{\Box} t_{rr'} = -1$) one obtains nodal Dirac fermions. We also study the projected wave function on square lattice with uniform hopping (and no flux).

The wave-functions for these spin-liquids are constructed by starting with a system of spin-$1/2$
fermionic spinons $f_{r\alpha}$ hopping on a finite lattice of size $L1 \times L2$ at half-filling with a mean field Hamiltonian:$
 H_{MF} = \sum_{r r'} \left[  -t_{r r'}f^{\dagger}_{r\sigma} f_{r'\sigma} + h.c. \right]$. The spin wave-function is given by $|\phi\rangle = P_G |\phi\rangle_{MF}$ where $|\phi\rangle_{MF}$ is the ground state of $H_{MF}$ and the Gutzwiller projector
$P_G = \prod_i \left( 1- n_{i \uparrow} n_{i \downarrow}\right) $ ensures exactly one fermion per site. The sign-structure of the projected wave-function depends markedly on the underlying lattice.
For a bipartite lattice with $t_{r r'}$ non-zero and real only for the opposite sublattices, one can prove that the wave-function satisfies the Marshall sign rule\cite{SupplementaryMaterial}.
Thus, for a bipartite lattice, one only needs to calculate
$\langle {\rm Swap}_{A,mod}\rangle$ since $\langle {\rm Swap}_{A,sign}\rangle$ trivially equals unity. The projected wave-function for the square lattice with and without $\pi$-flux (as well as that for the
one-dimensional Haldane-Shastry state) satisfies the Marshall's sign rule while that for the triangular lattice doesn't.
We discuss these three cases separately. The one dimensional case was previously discussed in \cite{Cirac}.

\begin{figure}[tb]
 \centerline{
   \includegraphics[width=200pt, height=200pt]{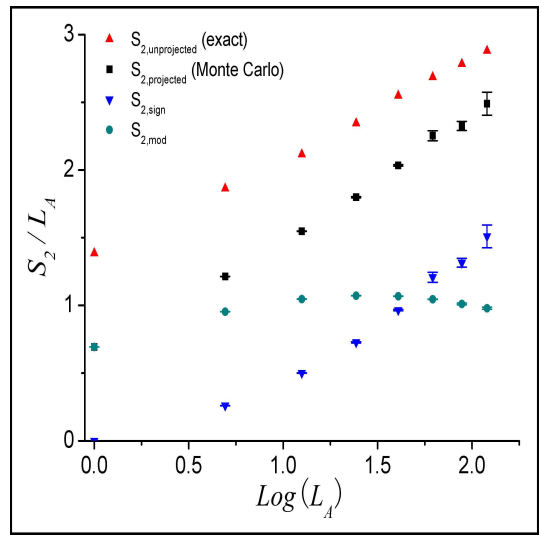}}
 \caption{Renyi entropy data for projected and unprojected Fermi sea state on the triangular lattice of size $18 \times 18$ with $L_A = 1\dots 8$. Note, projection barely modifies the slope, pointing to a Fermi surface surviving in the spin wavefunction. We also separately plot $S_{2,sign}$ and $S_{2,mod}$ (as defined in the text) for the projected state, the former dominates at larger sizes.}
 \label{tri}
 \end{figure}

\textit{Triangular lattice}:
As mentioned above the mean-field ansatz describes a spin-liquid with spinons hopping on a triangular lattice. We consider a lattice with total size $18 \times 18$
on a torus and the region $A$ of square geometry with linear size $L_A$ upto 8 sites. We find a clear signature of $L_A \,\textrm{log} L_A$ scaling in Renyi entropy (Fig. \ref{tri}).
This is rather striking since the wave-function is a spin wave-function and therefore could also be written in terms of hard-core bosons. This result strongly suggests the presence of an underlying spinon Fermi surface. In fact the coefficient of the $L_A \,\textrm{log} L_A$ term is rather similar before and after projection. This observation may be rationalized by picturing a two dimensional Fermi surface as a collection of many  independent one dimensional systems in momentum space, each giving rise to a $\log L$ contribution. Gutzwiller projection then just removes a single charge degree of freedom.

 \begin{figure}[tb]
 \centerline{
   \includegraphics[width=200pt, height=200pt]{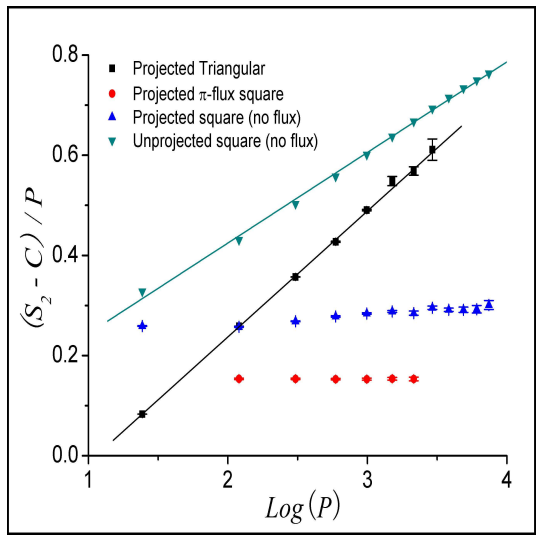}}
\caption{Renyi entropy for the projected Fermi sea state on the triangular lattice and square (with and without $\pi$-flux) lattice as a function of the perimeter $P$ of the subsystem $A$. Here $C$ is the constant part of the $S_2$. We find $S_2 \sim P log(P) + C$ for the projected triangular lattice state while $S_2 \sim P + C$ for the projected $\pi$-flux square lattice state. For the square lattice state (no flux), the projection leads to a significant reduction in $S_2$ suggesting at most a very weak violation of the area-law.}
 \label{all}
 \end{figure}


It is interesting to compare the contribution to $S_2$ from  $S_{2,sign} \equiv -\log(\left <{\rm Swap}_{A,sign}\right >) $ and  $S_{2,mod} \equiv -\log(\left <{\rm Swap}_{A,mod}\right >)$ separately.
Numerically, $S_{2,sign}$ appears to be responsible for the logarithmic violation of the area law  (Fig. \ref{tri}). This suggests that the sign structure of the wavefunction is crucial at least in this case.

The area-law violation of the Renyi entropy
for Gutzwiller projected wave-functions substantiates the theoretical expectation that an underlying Fermi surface is present in the spin wavefunction.

\textit{Square lattice with $\pi$ flux}:
The mean-field ansatz consists of spinons with Dirac dispersion around two nodes, say, $(\pi/2,\pi/2)$ and $(\pi/2,-\pi/2)$ (the locations of the nodes depend on the
gauge one uses to enforce the $\pi$ flux). The projected wave-function has been proposed in the past as the ground state of an algebraic spin liquid. The algebraic spin-liquid
is believed to be describable by a strongly coupled conformal field theory of Dirac spinons coupled to a non-compact $SU(2)$ gauge field \cite{wen2002, wen2004}. Because of this
the algebraic spin-liquid has algebraically decaying spin-spin correlations. We verify this explicitly for the projected wavefunction using Variational Monte Carlo on a $36 \times 36 $ lattice
\cite{SupplementaryMaterial}. This state is different from that in Ref. \cite{yao2010}, where Majorana fermions are coupled to a discrete $Z_2$ gauge field making them effectively free at low energies, in contrast to our critical state.

Square lattice being bipartite, the projected wavefunction satisfies Marshall's sign rule and hence we were able to perform Monte Carlo calculation of Renyi entropy on bigger lattice sizes
compared to the triangular lattice case. We chose the overall geometry as a torus of size $L_A \times 4 L_A$ with both region $A$ and its complement of sizes  $L_A \times 2 L_A$
(the total boundary size being $L_A + L_A = 2 L_A$). We considered $L_A$ upto 14 sites . We found that the projected wavefunction follows an area law akin to its unprojected counterpart
and has the scaling $S_2 \approx 0.30 L_A  + g$ where $g \approx 1.13$ is a universal constant that depends only on the aspect ratio of the geometry \cite{ryu2006}.
This is consistent with the presence of Dirac fermions at low energies.


\textit{Square lattice without any flux}:
The unprojected Fermi surface is nested. Since projection amounts to taking correlations into account, one might wonder whether the Fermi surface undergoes a magnetic instability after the projection.
 Indeed, we found non-zero magnetic order in the projected wave-function, consistent with an independent recent study Ref. \cite{li2011}. This was verified by calculating spin-spin correlations
 on a $42 \times 42$ lattice \cite{SupplementaryMaterial}. Renyi entropy calculations were done on a lattice of total size $24 \times 24$ with region $A$ being a square upto size $12 \times 12$.
The results are shown in the Fig. \ref{all}.
Though it is difficult to rule out presence of a partial Fermi surface from Renyi entropy, there is a significant reduction in the Renyi entropy as well as the coefficient of $L_A log L_A$ term
as compared to the unprojected Fermi sea.


{\em Summary:} In this paper we described a route to calculating the Renyi entropy $S_2$ for a wide variety of wave-functions that can handled by the  Variational Monte Carlo method. Our starting point
 is the ground state wave-function rather than the Hamiltonian. This allows us to study interesting  states on relatively  large systems, that may not be obtainable from sign problem free Hamiltonians.  We calculated
$S_2$ for Gutzwiller projected states that have been conjectured as the ground state wave-functions for gapless spin-liquids. We found that the projected Fermi sea on the triangular lattice violates the area law strongly indicating the presence of an emergent Fermi surface of neutral fermions. The sign structure of the wave-function makes a dominant contribution to the entropy.
 We note our algorithm is readily generalizable to calculate the higher Renyi entropies $S_n$ ($n > 2$). Also, it can be applied to study partially projected Fermi sea wave-function which model correlated Fermi liquids. This opens a window to study the entanglement entropy of fermions away from the free limit. A very interesting direction that we leave for the future, is testing the Widom conjecture, proposed currently for free fermions\cite{Klich}, to interacting states with Fermi surfaces.

{\em Acknowledgements:} We thank M. Fisher, A. Kallin, M.
Levin, R. Melko, O. Motrunich, M. Oshikawa, B. Swingle,
C. Xu for helpful discussions, and NSF DMR- 0645691 and NSF PHY05-51164 for support.

\end{document}